\documentclass[a4paper,twocolumn]{revtex4-1}
\pdfoutput=1 

\usepackage[T1]{fontenc}
\usepackage{graphicx}
\usepackage{amsmath}
\usepackage{amssymb}
\usepackage{amsthm}
\usepackage{dcolumn}
\usepackage{epsfig}
\usepackage{anysize}
\usepackage{bm}
\usepackage{datetime}
\usepackage[title]{appendix}
\usepackage[dvipsnames]{xcolor}
\usepackage{threeparttable}
\usepackage{textcomp}
\usepackage{booktabs}
\usepackage[paperwidth=210mm,paperheight=297mm,centering,hmargin=2cm,vmargin=2.5cm]{geometry}

\begin{document}

\title{Contour forward flux sampling: Sampling rare events \\along multiple collective variables}
\author{Ryan S. DeFever}
\author{Sapna Sarupria}
\email{ssarupr@g.clemson.edu}
\affiliation{Department of Chemical \& Biomolecular Engineering \\ Clemson University, Clemson, SC 29634}

\begin{abstract}
Many rare event transitions involve multiple collective variables
(CVs) and the most appropriate combination of CVs is generally unknown
a priori. We thus introduce a new method, contour forward flux
sampling (cFFS), to study rare events with multiple CVs
simultaneously. cFFS places nonlinear interfaces on-the-fly from the
collective progress of the simulations, without any prior knowledge of
the energy landscape or appropriate combination of CVs. We demonstrate
cFFS on analytical potential energy surfaces and a conformational
change in alanine dipeptide.
\end{abstract}

\maketitle

\section{Introduction}

Rare events remain uniquely challenging to study in molecular
simulations.\cite{PetersBook:17} These infrequent transitions between
long-lived (meta)stable states are characterized by large differences
between the timescales of the relevant physics (e.g., molecular
vibrations, hydrogen bond lifetimes, etc.) and the time between events
(often $\mu$s to s). Exemplars include crystal
nucleation\cite{Li:14:JPCB,
  Debenedetti:14:PCCP,Debenedetti:15:PNAS,DeFever:17:JCP}, ion-pair
dissociation in solution,\cite{Chandler:99:JPCB,Mundy:15:JPCB}
conformational changes in
biomolecules\cite{Bolhuis:06:PNAS,Escobedo:10:JCP}, and chemical
reactions\cite{vanErp:18:PNAS}. Due to the prevalence and importance
of rare events, several advanced sampling methods have been
developed\cite{Bennett:77:AlgChemComput,Chandler:98:JCP,Vanden-Eijnden:02:PRB,Bolhuis:03:JCP,Allen:05:PRL,Allen:06:JCP,Ciccotti:06:JCP,
  vanErp:07:PRL,Parrinello:13:PRL,Hummer:17:JCP,Dellago:10:RevCompChem}
to estimate transition rate constants and sample unbiased trajectories
connecting the stable states. However, even with increasing
computational power some phenomena remain challenging to study and
continued method development is required.

We present contour forward flux sampling (cFFS), a novel method to
sample rare events with multiple collective variables\footnote[2]{In
  this work, a {\it collective variable} is a quantity that can be
  calculated from the configuration space coordinates of the
  system. An {\it{order parameter}} is a collective variable that can
  distinguish between states $A$ and $B$.} (CVs)
simultaneously. Building on forward flux sampling (FFS), cFFS
leverages overall trajectory behavior to on-the-fly determine
nonlinear interface placement in multiple CVs. FFS is a rare event
sampling method that uses a series of non-overlapping interfaces to
drive a system from an initial state $A$ to final state
$B$.\cite{Allen:05:PRL,Allen:06:JCP,Allen:09:JPhys,Escobedo:09:JPhysCondMat}
Each interface is defined by some value of an order parameter,
$\lambda$, which changes monotonically from $A$ to
$B$. Straightforward simulation in $A$ is used to estimate the flux,
$\Phi_{A0}$, from $A$ to the first interface, $\lambda_0$, and to
collect a large number of first-crossing phase points at
$\lambda_0$. The designation of a phase point as a first-crossing
  point indicates that upon following the trajectory backwards in time
  from the point, one would reach $\lambda_A$ before $\lambda >
  \lambda_0$. Several trajectories are initiated from each phase
point collected at $\lambda_0$ ($\lambda_i$). Stochasticity from the
dynamics or velocity perturbation at the start of each simulation
ensures trajectory divergence. Trajectories returning to $A$ are
discarded, while those reaching the next interface, $\lambda_1$
($\lambda_{i+1}$), are stored for the next iteration. This procedure
is repeated for each interface until the boundary of $B$ is reached,
or the probability of advancing to the next interface,
$P(\lambda_{i+1}|\lambda_i)$, plateaus to 1. The transition rate
constant is calculated as $k_{AB} =
\Phi_{A0}\prod_{i=0}^{n-1}P(\lambda_{i+1}|\lambda_i)$ and transition
paths from $A$ to $B$ are generated by connecting the partial paths
backward from $B$ to $A$. FFS has emerged as a popular choice for
studying rare events in simulation because it is applicable to
equilibrium and nonequilibrium systems, and implementation is
algorithmically straightforward and embarrassingly parallel.

Despite it's advantages, FFS has shortcomings. Assuming
  reasonable definitions for the boundaries of $A$ and $B$, the rate
  constant and transition path ensemble (TPE) computed with FFS are,
  in principle, independent of the order parameter used for the
  calculation.\cite{Allen:06:JCP} In practice, a poor choice of order
  parameter is detrimental to the efficiency of
  FFS\cite{vanErp:06:JCP,vanErp:12:AdvChemPhys} and can even lead to
  incorrect results.\cite{Bolhuis:08:Biophys,vanErp:12:AdvChemPhys}
  This arises when portions of $\lambda_i$ which are important to the
  transition are sparingly sampled. More formally, imagine some
  coordinate ($\lambda^\perp$) orthogonal to $\lambda$. Challenges
  arise for FFS when there is poor overlap between the distribution of
  first-crossing phase points captured at $\lambda_i$,
  $\rho(\lambda^\perp | \lambda_i)$, and the probability of reaching
  $\lambda_B$ from some point on $\lambda_i$, $P(\lambda_B |
  \lambda_i; \lambda^\perp)$.\cite{vanErp:12:AdvChemPhys} There are
  two approaches to overcome this issue: (1) increase sampling to
  collect more phase points at problematic interface(s), or (2)
  improve the choice of order parameter to increase overlap between
  the two distributions. The first approach yields more phase points
  everywhere along an interface, but with sufficient sampling the
  paths spawned from phase points with a higher $P(\lambda_B |
  \lambda_i; \lambda^\perp)$ will come to dominate the eventual path
  ensemble, resulting in the correct rate constant and
  TPE. Unfortunately, the efficiency of FFS will still be poor. In
  contrast, the second approach increases the efficiency of FFS,
  meaning that FFS will converge to the correct rate constant and TPE
  with less sampling. Unfortunately, optimal order parameters are
  rarely known a priori. More often, one of the reasons for generating
  a path ensemble with a method such as FFS is to identify order
  parameters which best describe the transition.

Since sampling of all interfaces $i>0$ in FFS depends on the
  phase points collected at $\lambda_0$, methods have been proposed to
  optimize placement of, and ensure adequate sampling of
  $\lambda_0$.\cite{Escobedo:09:JCP,Li:14:JPCB,Michaelides:16:JPCL} If
  the situation is not too dire, increasing the length of the basin
  simulation and collecting more phase points at $\lambda_0$ may
  provide a sufficient remedy. However, if overlap between
  $\rho(\lambda^\perp | \lambda_0)$ and $P(\lambda_B | \lambda_0;
  \lambda^\perp)$ is extremely small, this may be
  insufficient. Furthermore, the problem is not limited to
  $\lambda_0$; in principle the distribution of phase points sampled
  at any $\lambda_i$ could suffer from this problem. Poor overlap
  between $\rho(\lambda^\perp | \lambda_i)$ and $P(\lambda_B |
  \lambda_i; \lambda^\perp)$ becomes particularly problematic for
  systems with multiple transition tubes. There, a poor choice of
  order parameter may result in some transition tubes becoming
  (artificially) favored over others. In the extreme, entire
  transition tubes can be missed by FFS.

A related situation worth mentioning is when $\rho(\lambda^\perp
  | \lambda_0)$ converges extremely
  slowly.\cite{Li:14:JPCB,Michaelides:16:JPCL} If this is the problem,
  extending the basin simulations until convergence is achieved will
  remedy the situation.\cite{Li:14:JPCB} A greater number of phase
  points at $\lambda_0$ are not required; just phase points correctly
  sampled from the converged distribution.

The choice of order parameter strongly affects the overlap
  between $\rho(\lambda^\perp | \lambda_i)$ and $P(\lambda_B |
  \lambda_i ; \lambda^\perp)$. If the order parameter is the committor
  function, $P(\lambda_B | \lambda_i ; \lambda^\perp)$ is constant
  with $\lambda^\perp$, thereby assuring good overlap between
  $\rho(\lambda^\perp | \lambda_i)$ and $P(\lambda_B | \lambda_i ;
  \lambda^\perp)$.\cite{vanErp:12:AdvChemPhys} Borrero and Escobedo
thus devised a method to optimize the order parameter with a series of
FFS simulations.\cite{Escobedo:07:JCP} Though the approach yields
improvements\cite{Escobedo:10:JCP}, it is challenging for systems
which require extraordinary computational resources for even a single
FFS run.\cite{Debenedetti:15:PNAS,DeFever:17:JCP} Furthermore, some
processes are inherently
multidimensional,\cite{Tanaka:16:JCP,Clementi:13:AnnRev,Debenedetti:15:JCP,Chandler:99:JPCB}
and driving the transition along a single CV may not be ideal.

cFFS takes a different approach. We extend FFS to use multiple
  CVs on-the-fly. This allows researchers to test multiple CVs
  simultaneously and improves the chances of capturing important
  orthogonal coordinates within the set of CVs used to drive the
  transition. At each interface, cFFS identifies the next interface
as a nonlinear combination of specified CVs on-the-fly from the
behavior of simulations initiated from the previous interface. In
doing so, cFFS also reveals the role of each CV through the entire
transition. Only the combination of CVs must separate $A$ and $B$ and
so each CV need not monotonically change from $A$ to $B$. If some CV
is unimportant, this will be reflected by, but not impede cFFS. These
features offer substantial flexibility in CVs that can be used with
cFFS. cFFS generates an estimate of the transition rate constant and a
collection of $A \rightarrow B$ trajectories belonging to the TPE. We
demonstrate cFFS with two CVs, but in principle it can be extended to
three or more CVs.

In Sec. II we explain cFFS. We proceed to demonstrate cFFS on several
two-dimensional potential energy surfaces in Sec. III. In Sec. IV, we
demonstrate cFFS with one position coordinate and one momentum
coordinate, and in Sec. V we test cFFS on a standard higher
dimensional test case, a conformational transition in alanine
dipeptide. Discussion and closing remarks are provided in Sec. VI and
Sec. VII, respectively.

\section{Contour forward flux sampling}

The central idea of cFFS is to allow the system to naturally
  evolve along multiple CVs to reveal how different CVs participate in
  the transition. This is achieved by placing the subsequent interface
  based on sampling initiated from the current interface. The FFS
  formalism can still be used to calculate the rate constant and
  TPE. Interface placement is designed such that the distribution of
  first-crossing points is uniform along the interface, ensuring that
  each interface is well-sampled everywhere within the chosen CVs.

The first step of cFFS is to run straightforward basin simulations in
$A$ to identify the bounds of $A$ ($\lambda_A$) and the first
interface ($\lambda_0$), and to collect phase points at
$\lambda_0$. The value of each CV in time, $\bm{\lambda}(t)$, is
calculated, where $\bm{\lambda} \equiv \{ \lambda^I, \lambda^{II},
\dots, \lambda^N \}$ is the set of CVs. CV space is discretized to
create an $N$-D grid. The discretization size is selected such that
the system rarely travels more than a single grid site in one time
step. The discrete probability distribution, $P(\bm{\lambda})$, is
calculated from the basin simulations. Grid sites exceeding a
threshold probability are added to the set of sites describing $A$,
$\bm{s}_A$. Regions of CV space which are not in $\bm{s}_A$ but
completely surrounded by $\bm{s}_A$ are added to
$\bm{s}_A$. $\lambda_A$ is defined as the boundary between sites in
$\bm{s}_A$ and those that are not. Trajectories exit $A$ when they
cross from a grid site in $\bm{s}_A$ to a grid site not in $\bm{s}_A$.

Several criteria are used to identify $\bm{s}_0$, the set defining
$\lambda_0$. $\bm{s}_0$ should: (a) completely contain $\bm{s}_A$ so
that $\lambda_0$ does not overlap with or cross $\lambda_A$, (b) not
create regions of CV space completely surrounded by $\bm{s}_0$, but
not included in it, (c) not include sites in $\bm{s}_B$, the set of
sites describing $B$, (d) be selected such that some desired number of
phase points can be collected at $\lambda_0$, and (e) be selected such
that there is equal flux of trajectories exiting $\bm{s}_0$ along the
entire $\lambda_0$ interface. Criteria (e) is crucial as it ensures
that cFFS does not bias the system to sample any one direction more
readily than another. Further discussion is provided later. Once
$\lambda_A$ and $\lambda_0$ are defined the basin simulations are
re-analyzed to calculate $\Phi_{A0}$ and collect phase points at
$\lambda_0$.

The remainder of cFFS proceeds as follows. Several trajectories are
initiated from each phase point at $\lambda_i$ ($\lambda_i=\lambda_0$
for the first iteration). Trajectories are terminated when they return
to $\lambda_A$, or reach a maximum number of steps. The set of sites
defining $\lambda_{i+1}$, $\bm{s}_{i+1}$, is determined from the
behavior of trajectories initiated at $\lambda_{i}$ using analogous
criteria to those described for determining $\lambda_0$. Note that
$\bm{s}_{i+1}$ must completely contain $\bm{s}_{i}$ to satisfy the
effective positive flux
formalism.\cite{Bolhuis:03:JCP,Bolhuis:05:JComputPhys} Once
$\bm{s}_{i+1}$ is identified, trajectories are re-analyzed to
determine if they cross $\lambda_{i+1}$ (i.e., exit $\bm{s}_{i+1}$)
before returning to $A$. For each trajectory that crosses
$\lambda_{i+1}$, the phase point at the time step which the trajectory
crosses $\lambda_{i+1}$ is saved. Trajectories which fail to reach
$\lambda_{i+1}$ or return to $A$ before the maximum number of
    steps are extended until they reach $\lambda_{i+1}$ or
    return to $A$. The probability, $P(\lambda_{i+1} |
\lambda_i)$, is calculated from the number of trajectories that reach
$\lambda_{i+1}$ before returning to $A$.

Eventually, sites in $\bm{s}_{i+1}$ will be adjacent to sites in
$\bm{s}_B$. Trajectories initiated from $\lambda_i$ can then reach
$\lambda_{i+1}$, return to $A$, or proceed directly to $B$. This
indicates the kinetic barrier has been surmounted and thus cFFS is
nearly complete. Two probabilities are now calculated;
$P(\lambda_{i+1} | \lambda_i)$ and $P(\lambda_B | \lambda_i)$. Our
approach is to continue cFFS until $\bm{s}_{i+1}$ surrounds
$\bm{s}_B$. At this point, $i$ becomes the final interface,
$n$. Trajectories initiated from $\lambda_{n}$ are continued until
they reach $\lambda_B$ or return to $\lambda_A$ to close the
probabilities for the rate calculation. As with multi-state
FFS\cite{Bolhuis:18:JCP}, the transition rate constant is calculated
as

\begin{equation}
k_{AB} = \Phi_{A0} \sum_{j=0}^{n} P(\lambda_B | \lambda_{j}) \prod_{i=0}^{j-1} P(\lambda_{i+1} | \lambda_i ). 
\end{equation}

The collection of trajectories comprising the TPE is constructed by
connecting the partial paths backwards from $B$ to $A$. Note that all
trajectories {\it do not} have equal weight in the TPE. The relative
weight of each trajectory is $w=1/\prod_{i=0}^j k_i$, where $j$ is the
final interface crossed by a trajectory before reaching $B$ and $k_i$
is the number of trajectories initiated from each configuration at
interface $i$.

\section{Demonstration on 2D potential energy surfaces}

\begin{figure*}[t]
  \vspace{-0.2in}
    \centering
    \includegraphics[width=0.85\linewidth]{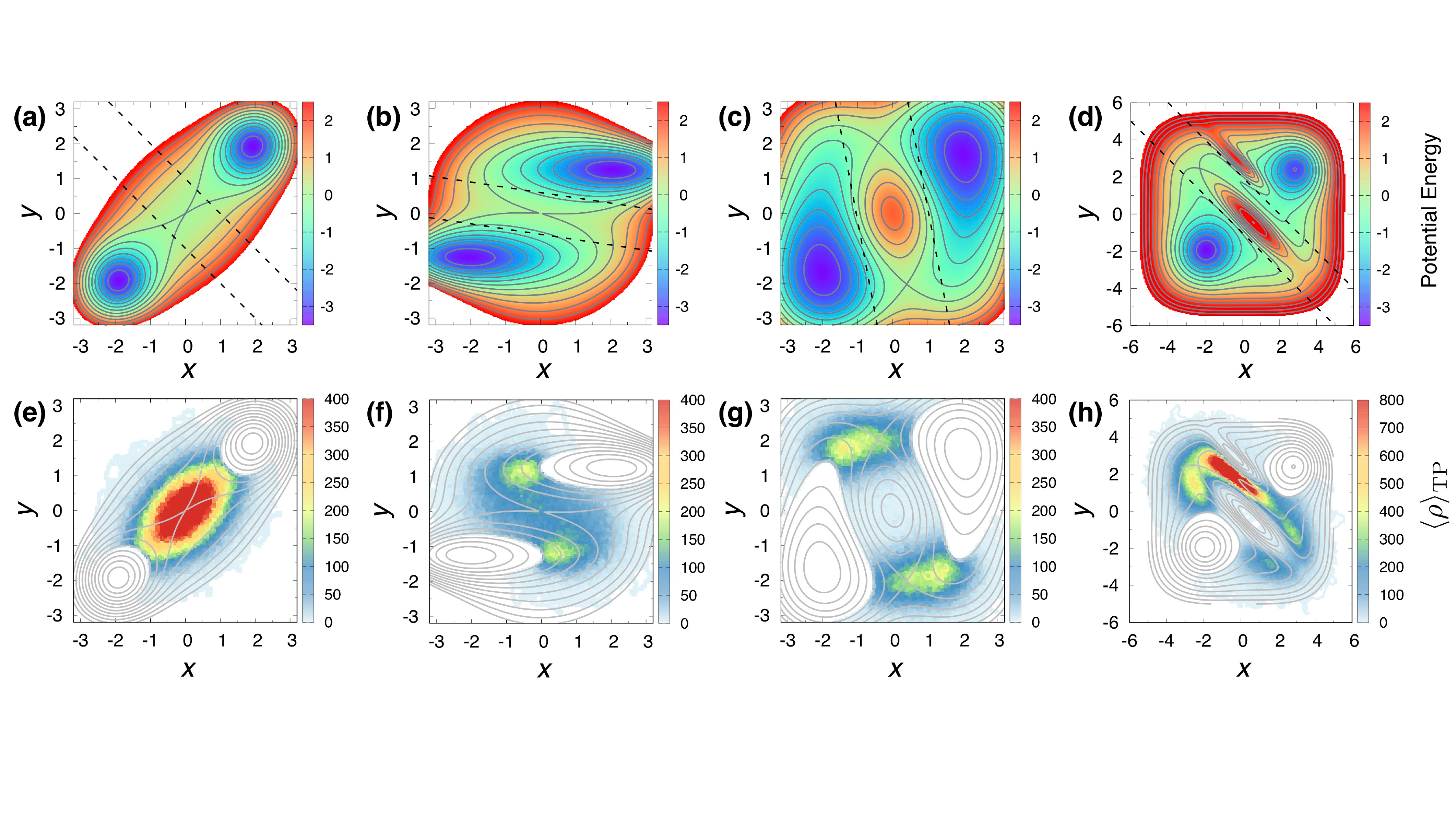}
    \vspace{-0.6in}
    \caption{{\it Top panels:} PESs used to test cFFS: (a) PES-1, (b)
      PES-2, (c) PES-3, and (d) PES-4. Color represents the potential
      energy. Contour lines are separated by 0.5 units. The region
      between the dashed lines was used to quantitatively compare
      $\rho(q | \text{TP})$ between different methods. {\it Bottom
        panels}: TPE sampling from SLD at $\beta=2.5$ on (e) PES-1,
      (f) PES-2, (g) PES-3, and (h) PES-4.}
    \label{fig:PESs}
    \vspace{-0.12in}
\end{figure*}

We demonstrate cFFS with Langevin dynamics of a single particle on
four 2D potential energy surfaces (PESs) with different topographical
features (see Fig. \ref{fig:PESs}(a)--(d)). PES-1 has a single
    transition tube which follows two monotonically increasing
    CVs. PES-2 has a single transition tube with hysteresis in the $x$
    coordinate. PES-3 and PES-4 both contain two transition tubes; the
    potential energy barriers are the same for the two tubes on PES-3, and
    different for the two tubes on PES-4. For each PES, we study
$A\rightarrow B$ transitions with straightforward Langevin dynamics
(SLD), FFS$_\text{opt}$, FFS$_\text{x}$, and cFFS. FFS$_\text{opt}$
denotes FFS performed with the optimal linear combination of $x$
    and $y$ (i.e., the order parameter orthogonal to the dividing
surface of the PES), and FFS$_\text{x}$ indicates FFS performed with
$x$ as the (suboptimal) order parameter. We stress that optimal order
parameters are not known a priori for most realistic systems, and
therefore FFS is generally performed with suboptimal order
parameters. Further details of the PESs, Langevin dynamics, and
FFS/cFFS parameters are provided in the Supporting Information (SI).

\subsection{Rate constants}

\begin{table}[h]
  \vspace{-0.01in} \linespread{0.9}\selectfont\centering
  \begin{center}
\begin{threeparttable}
\caption{$A \rightarrow B$ transition rate constants for four 2D
  PESs. One standard deviation of the mean is reported in
  parenthesis.}
\label{tab:rates}
\setlength{\tabcolsep}{6pt}

\begin{tabular}{c c c c c}
  \toprule
  & \multicolumn{4}{c}{$k_{AB} \times 10^5$ at $\beta=2.5$} \\ 
  \cmidrule{2-5}
PES & SLD & FFS$_\text{opt}$ & FFS$_\text{x}$  & cFFS  \\
\midrule
PES-1 & 2.9 (0.2)  & 2.8 (0.3)  & 3.1  (0.9)   & 2.8 (0.2) \\
PES-2 & 9.1 (0.1)  & 7.9 (0.9)  & 10.2 (2.5)   & 8.8 (0.7) \\
PES-3 & 2.6 (0.3)  & 2.4 (0.4)  & 2.3  (0.6)   & 2.4 (0.1) \\
PES-4 & 1.1 (0.1)  & 1.0 (0.1)  & 1.1  (0.1)   & 1.0 (0.1) \\
\midrule
& \multicolumn{4}{c}{$k_{AB} \times 10^9$ at $\beta=5.0$} \\
\cmidrule{2-5}
PES-1 & 5.4  (1.2) & 4.4  (0.2)  & 3.1  (0.2)  & 4.5 (0.6) \\
PES-2 & 23.2 (2.0) & 18.0 (3.5)  & 18.3 (1.0)  & 21.9 (2.3) \\
PES-3 & 6.4  (1.3) & 2.9  (0.2)  & 2.5  (0.2)  & 5.4 (0.5) \\
PES-4 & 2.8  (0.9) & 1.9  (0.4)  & 0.42 (0.02) & 2.6 (0.1) \\
\bottomrule
\end{tabular}
\end{threeparttable}
  \end{center}
  \vspace{-0.1in}
\end{table}

$A \rightarrow B$ transition rate constants are reported in Table
\ref{tab:rates}. Transitions were studied at $\beta=2.5$ and
$\beta=5.0$ ($\beta=1/k_BT$). The higher temperature ($\beta=2.5$)
enables rigorous comparison of TPE sampling with SLD, whereas the
lower temperature ($\beta=5.0$) provides a test at more challenging
conditions. SLD rate constants are unbiased estimates. FFS$_\text{x}$
provides accurate estimates of the rate constants at $\beta=2.5$, but
at $\beta=5.0$ FFS$_\text{x}$ underestimates the rate constants. This
suggests that suboptimal order parameters perform worse as the barrier
becomes larger relative to $k_BT$. We explain the breakdown of
FFS$_\text{x}$ by examining the TPE sampling below. FFS$_\text{opt}$
and cFFS perform better. Rate constants from FFS$_\text{opt}$ and cFFS
both agree nicely with SLD at $\beta=2.5$. At $\beta=5.0$,
FFS$_\text{opt}$ underestimates rate constants for PES-2 and
PES-3. In contrast, cFFS provides correct estimates of the
      rate constants for all four PESs at $\beta=5.0$.

\subsection{Transition path ensemble sampling}

Though attaining the correct $A \rightarrow B$ rate constant is a
crucial test of cFFS, it is also important that cFFS correctly samples
the TPE. TPE sampling is calculated as $\langle \rho \rangle_\text{TP}
= \langle n_\text{visits}/l^2 \rangle_\text{TP}$, where $\langle
... \rangle_\text{TP}$ indicates an ensemble average over all
transition paths, and $n_\text{visits}$ is the number of times a
transition path visited each $l\times l$ region of space. For
reference, TPE sampling from SLD at $\beta=2.5$ is shown in the bottom
panels of Fig. \ref{fig:PESs}. 

\begin{figure*}
    \centering
    \includegraphics[width=0.93\linewidth]{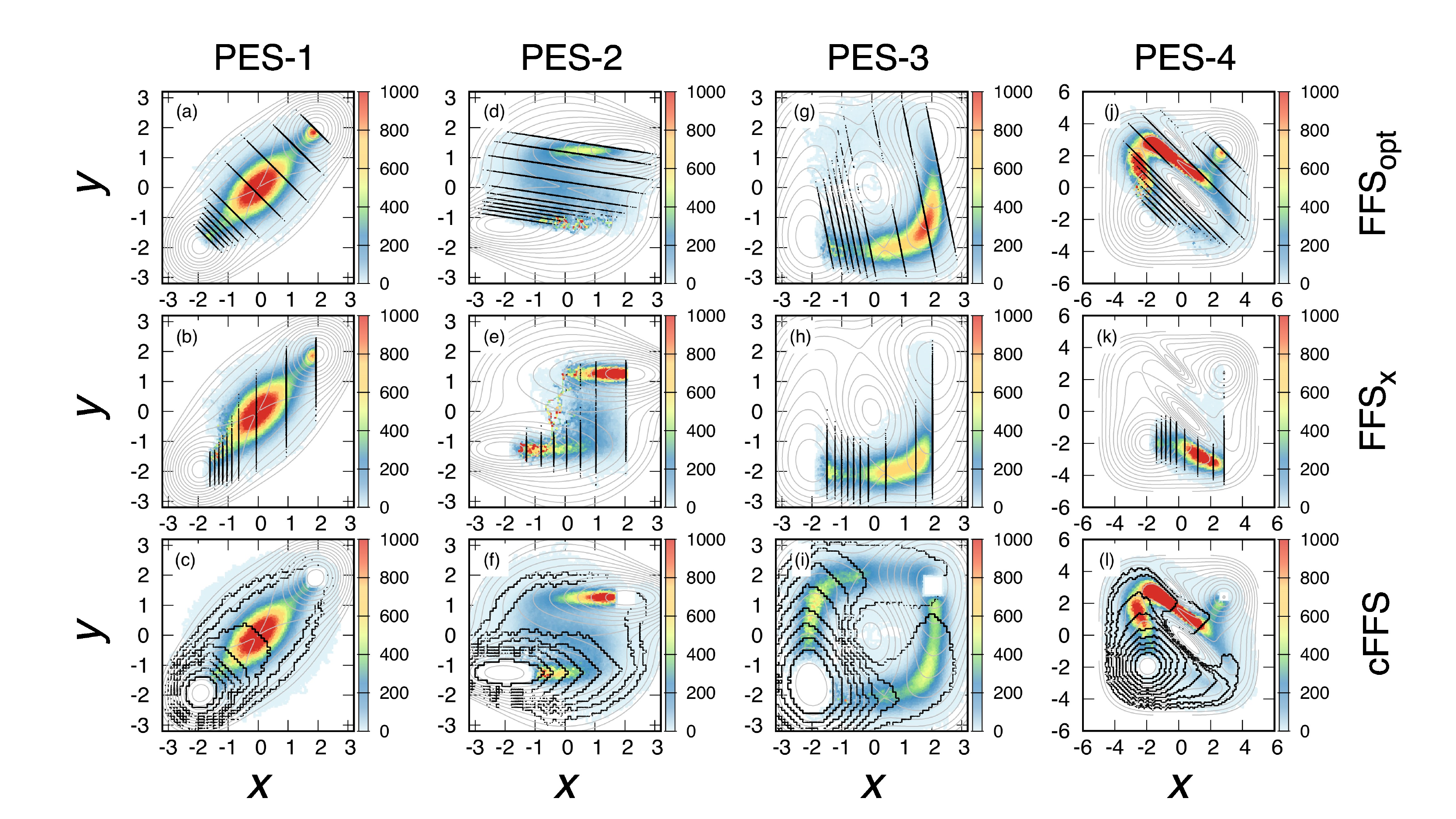}
    \vspace{-0.1in}
    \caption{Comparison of interface placement and TPE sampling
      generated with FFS$_\text{opt}$, FFS$_\text{x}$, and cFFS on
      PES-1 -- PES-4 at $\beta=5.0$. PES contours are shown as gray
      lines. Configurations collected at each interface are shown with
      black points. TPE sampling represented by the heat map.}
    \label{fig:tpe}
\end{figure*}

Fig. \ref{fig:tpe} summarizes the behavior of FFS$_\text{opt}$,
FFS$_\text{x}$, and cFFS on PES-1--PES-4 at $\beta=5.0$. All methods
result in qualitatively similar sampling for PES-1. The other
  surfaces proved more challenging for FFS$_\text{x}$ and
  FFS$_\text{opt}$. In contrast, cFFS results in the qualitatively
correct sampling for all four PESs. On PES-2, the hysteresis
  provides a challenge for FFS$_\text{x}$. Unlike FFS$_\text{opt}$ and
  cFFS, FFS$_\text{x}$ undersamples the $x<0$ portion of the
  transition tube. On PES-3 and PES-4, the failure of
  FFS$_\text{x}$ is even more stark; FFS$_\text{x}$ only samples one
of the two transition tubes. Even FFS$_\text{opt}$ fails to sample
both transition tubes equally on PES-3. PES-3 and PES-4 have two
distinct transition tubes, and the minimum energy paths change
direction from $A$ to $B$. On PES-3, both transition tubes have the
same potential energy barrier. However, one transition tube approaches
the transition state from $A$ with a gentler slope. Results from SLD
at $\beta=2.5$ in Fig. \ref{fig:PESs}(c) indicate that both
transitions should be equally traveled. cFFS reproduces this behavior
at both $\beta=2.5$ (SI Fig. S1) and the more challenging $\beta=5.0$
(Fig. \ref{fig:tpe}(i)). At $\beta=5.0$, FFS$_\text{x}$ only
samples a single transition tube (Fig. \ref{fig:tpe}(h)). Even
FFS$_\text{opt}$ struggles to sample both transition tubes equally on
PES-3 (Fig. \ref{fig:tpe}(g)). The behavior of FFS$_\text{opt}$
  and FFS$_\text{x}$ on PES-3 can be explained by the framework put
  forth in the introduction. In both cases, it is apparent that
  $\rho(\lambda^\perp | \lambda_0)$ sampled during the basin
  simulations only has good overlap with $P(\lambda_B | \lambda_0;
  \lambda^\perp)$ for one of the two transition tubes. The result is
  that FFS oversamples the tube with greater overlap, at the expense
  of the other transition tube. FFS sensitivity to the choice of order
parameter on PES-3 is further demonstrated in SI Fig. S2. Though
  FFS will converge to the correct TPE in the limit of infinite
  sampling, as a practical matter FFS can lead to incorrect results.
cFFS again performs well on PES-4, illustrating that cFFS is able to
navigate a tortuous transition landscape with two transition tubes and
unequal potential energy barriers.

\begin{figure}
    \centering
    \includegraphics[width=0.8\linewidth]{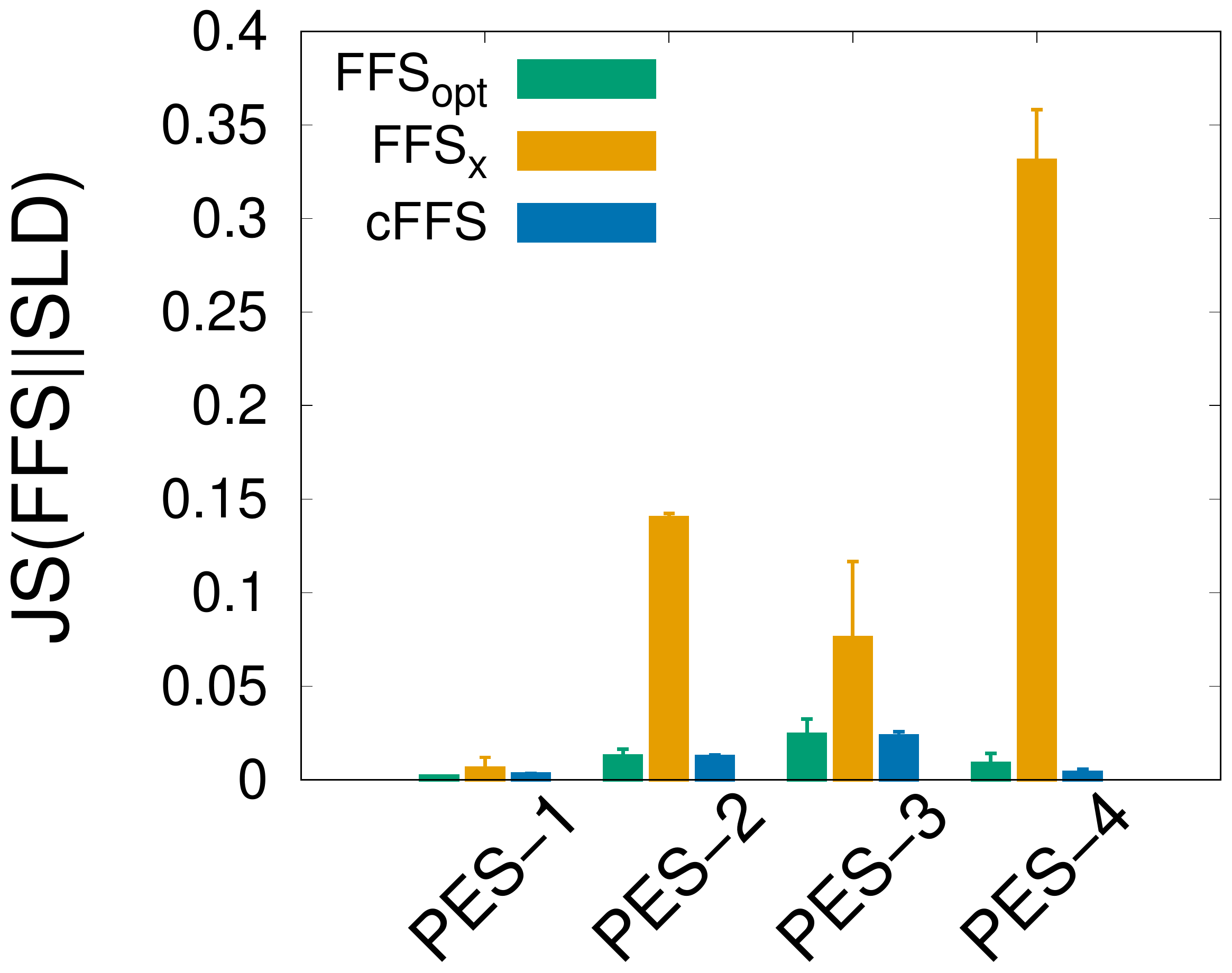}
    \caption{Jensen-Shannon divergence between $\rho(q | \text{TP})$
      calculated with SLD and FFS$_\text{opt}$, FFS$_\text{x}$, and
      cFFS at $\beta=2.5$. A value of zero indicates identical
      probability distributions, while a value of 1.0 indicates
      completely non-overlapping distributions. Error bars represent one standard
      deviation on the mean of three independent trials.}
    \label{fig:JS}
    \vspace{-0.1in}
\end{figure}

Near the dividing surface (see Fig. \ref{fig:PESs}) we quantitatively
compare the TPE density of states, $\rho(q|\text{TP})$, from SLD with
that from FFS$_\text{opt}$, FFS$_\text{x}$, and cFFS using the
Jensen-Shannon divergence\cite{Lin:91:IEEE}. We restrict our
comparison to $\beta=2.5$, where a large number of transitions can be
generated with SLD, hence providing a robust reference. The results
shown in Fig. \ref{fig:JS} confirm qualitative conclusions from
Fig. \ref{fig:tpe} ($\beta=5.0$) and SI Fig. S1 ($\beta=2.5$). At
$\beta=2.5$, FFS$_\text{opt}$ and cFFS perform similarly. For the
  simplest case (PES-1), FFS$_\text{x}$ performs nearly as well as
  FFS$_\text{opt}$ and cFFS. However, for the more complex surfaces,
  including the surface with hysteresis (PES-2), and surfaces with two
  transition tubes (PES-3, PES-4), FFS$_\text{x}$ performs notably
  worse.

\subsection{cFFS interface placement}

Fig. \ref{fig:tpe} also demonstrates cFFS interface
placement. Interfaces are spaced further apart in directions that
trajectories more readily advance and closer together in directions
that trajectories struggle to advance. For these low-dimensional
systems, interface locations adhere closely to the contours of the
PESs. We strongly emphasize that no knowledge of the PES is employed;
cFFS places interface $\lambda_{i+1}$ from the progress of
trajectories initiated from $\lambda_i$ alone.

If not done properly, performing FFS with multiple CVs simultaneously
can bias the system to over-sample or under-sample regions of CV
space. The amount of work performed by FFS is related to interface
spacing (i.e., $\lambda_{i+1}-\lambda_i$), slope of the free energy
landscape between $\lambda_i$ and $\lambda_{i+1}$, and the number of
trajectories initiated from $\lambda_i$. If the slope of the free
energy landscape between two interfaces becomes steeper,
$\lambda_{i+1}$ is moved closer to $\lambda_i$ or the number of
trajectories initiated from $\lambda_i$ is increased. Multiple CVs
introduces a new prospect; that unequal amounts of work are inserted
along different CVs, biasing the system to over-sample in the
direction that more work is inserted.

We introduced a condition of constant flux along an interface in cFFS
interface placement to address this problem. The force exerted by the
underlying free energy surface is proportional to
$-dn_\text{cross}/d\lambda$, where $n_\text{cross}$ is the number of
trajectories crossing an interface placed at some value of
$\lambda$. If $n_\text{cross}$ changes more quickly with changing
$\lambda$, then the underlying surface must have a steeper
slope. Applying the differential definition of work, $dW = Fd\lambda$,
and thus $dW \propto dn_\text{cross}$ and $W \propto
n_\text{cross}$. Constant flux along the interface requires that all
small sections of $\lambda_{i+1}$ have approximately the same number
of trajectories crossing them. This condition ensures that equal work
is inserted everywhere along the interface (i.e., in all directions)
and results in $\lambda_{i+1}$ closer to $\lambda_i$ in directions
trajectories struggle to advance and further from $\lambda_i$ in
directions trajectories readily advance. The fact that cFFS is able to
reproduce the correct TPE symmetry for PES-3 and PES-4 provides strong
evidence that the constant flux along the interface condition is
correct.

In complex systems, the optimal order parameter is often expected to
be a combination (linear or nonlinear) of multiple (suboptimal) order
parameters. This combination is generally nonintuitive and difficult
to predict. As such, most applications of FFS use a suboptimal order
parameter (e.g., FFS$_\text{x}$). On the four PESs, cFFS successfully
produces correct TPE sampling without knowing how $x$ and $y$ should
be combined. Though $x$ and $y$ are part of the optimal order
parameter, independently, $x$ and $y$ are suboptimal order
parameters. This suggests that cFFS can outperform FFS when multiple
suboptimal order parameters are known, but the optimal order parameter
remains unknown. In addition, nonlinear combinations of CVs have
increased degeneracy compared with linear combinations of CVs in
creating reaction coordinates (i.e., optimal order
parameters).\cite{Dellago:15:JPhysCondMat} Since cFFS interfaces are
arbitrarily complex combinations of the specified CVs, there may be
substantial flexibility in selecting good CVs for cFFS. A variety of
approaches have been proposed for identifying important CVs for rare
event
transitions.\cite{Hummer:05:PNAS,Dinner:05:JPCB,Peters:06:JCP,Coifman:09:PNAS,Bolhuis:10:JCP,Escobedo:07:JCP,Ma:14:MolSim,Peters:16:AnnRev}
For example, recent work suggests that important CVs can be identified
from local fluctuations in the (meta)stable
basins.\cite{Parrinello:18:JPCL} We envision using such approaches to
identify key CVs for cFFS.

\section{cFFS with a momentum coordinate}

FFS is most often applied in the diffusive limit and the CVs used
  as FFS order parameters are generally only functions of the atomic
  coordinates. In this section, we demonstrate cFFS on a simple
  analytical potential where momentum plays a key role during the
  transition. A previous study shows that FFS fails and under-predicts
  the transition rate constant when using a position-based order
  parameter alone.\cite{vanErp:12:AdvChemPhys}

Ref. \citenum{vanErp:12:AdvChemPhys} tested several path sampling
  methods for a transition on a simple 1D analytical potential
  described by $V(r) = r^4 - 2r^2$. Of the tested methods, replica
  exchange transition interface sampling (RETIS) and partial path
  transition interface sampling (PPTIS) provided the best estimates to
  the reference effective positive flux (EPF) rate ($k_{AB}^\text{EPF} = 2.4
  \pm 0.1 \times 10^{-7}$, $k_{AB}^\text{RETIS} = 2.8 \pm 0.7 \times
  10^{-7}$, $k_{AB}^\text{PPTIS} = 2.7 \pm 0.6 \times 10^{-7}$). FFS
  performed worst, underestimating the rate constant by 1--2 orders of
  magnitude depending on the length of the basin simulation. With a
  basin simulation of 4 million steps, FFS produced a rate constant of
  $k_{AB}^\text{FFS-short} = 2.2 \pm 0.2 \times 10^{-9}$. When the
  basin simulation was extended to 10 million steps,
  $k_{AB}^\text{FFS-long} = 1.2 \pm 0.1 \times 10^{-8}$. As explained
  in Ref. \citenum{vanErp:12:AdvChemPhys}, the source of systematic
  error in the rate constant was the lack of overlap between
  $\rho(\lambda^\perp | \lambda_0)$ and $P(\lambda_B | \lambda_0;
  \lambda^\perp)$. Successful transitions require large momentum when
  exiting the initial basin, and few to none of the trajectories
  captured at $\lambda_0$ had the requisite momentum. Even successful
  transition paths from FFS exited the initial state with lower
  momenta compared with other methods, resulting in a low estimate of
  the rate constant. This also resulted in the unphysical result that
  the momenta of transition paths from FFS were not symmetric about
  the barrier.

\begin{figure}
    \centering
    \includegraphics[width=0.8\linewidth]{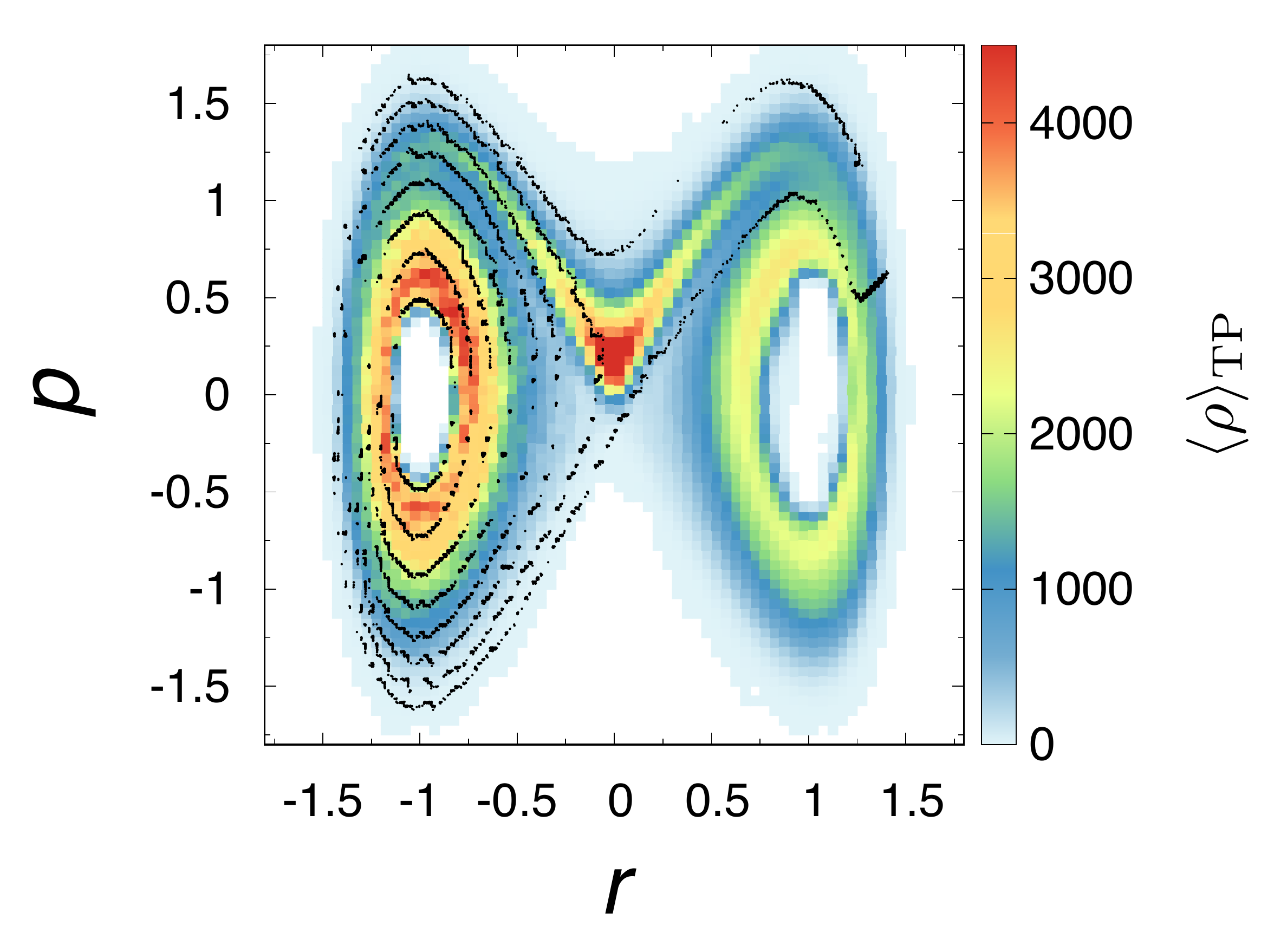}
    \caption{cFFS on 1D potential with one position coordinate ($r$) and
      one momentum coordinate ($p$). Initial basin $A$ is $r < 0$ minima
      and final basin $B$ is $r>0$ minima. Configurations collected at
      each interface are shown as black points. Color map shows the TPE
      sampling.}
    \label{fig:1dtpe}
     \vspace{-0.12in}
\end{figure}

We perform cFFS with the above potential at identical conditions
  as Ref. \citenum{vanErp:12:AdvChemPhys}. The two variables for cFFS
  are the position ($r$) and momenta ($p$). The basin simulation is
  performed with 4 million steps. We place interfaces adaptively,
  collecting $\sim$2,000 configurations per interface. As in
  Ref. \citenum{vanErp:12:AdvChemPhys}, we initiate 20,000
  trajectories from each interface. cFFS resulted in shooting from 8
  interfaces, compared with the 7 interfaces used in
  Ref. \citenum{vanErp:12:AdvChemPhys}. The average rate constant from
  three cFFS trials was $k_{AB}^\text{cFFS} = 2.0 \pm 0.1 \times
  10^{-7}$, slightly underestimating the EPF rate constant from
  Ref. \citenum{vanErp:12:AdvChemPhys}. The TPE and configurations
  collected at each interface from cFFS are shown in
  Fig. \ref{fig:1dtpe}. Paths exit the initial state orbiting the
  basin and acquiring more kinetic energy until they are able to
  escape. Their momenta then approaches zero as they cross through the
  transition state, before accelerating towards and orbiting into the
  final state. Consistent with theoretical expectations, the TPE
  generated by cFFS is symmetric about the barrier. We also tested
  cFFS with less sampling. Even with a twenty-fold reduction in
  sampling (1000 trajectories, 100 configurations per interface), the
  rate constant calculated with cFFS is $k_{AB}^\text{cFFS} = 2.6 \pm
  0.7 \times 10^{-7}$ and the TPE remains symmetric about the
  barrier.

These results demonstrate the potential for using cFFS to study
  transitions with important momenta variables. Though the above test
  case represents an extremely simple analytical model, it
  demonstrates the advantages of cFFS in such scenarios. If an
  important momenta variable is known for a transition, cFFS allows
  the basins to be separated with a position coordinate and the
  momentum coordinate can be used to help drive the transition.

\section{Demonstration on alanine dipeptide}

\begin{figure}
    \centering
    \includegraphics[width=0.8\linewidth]{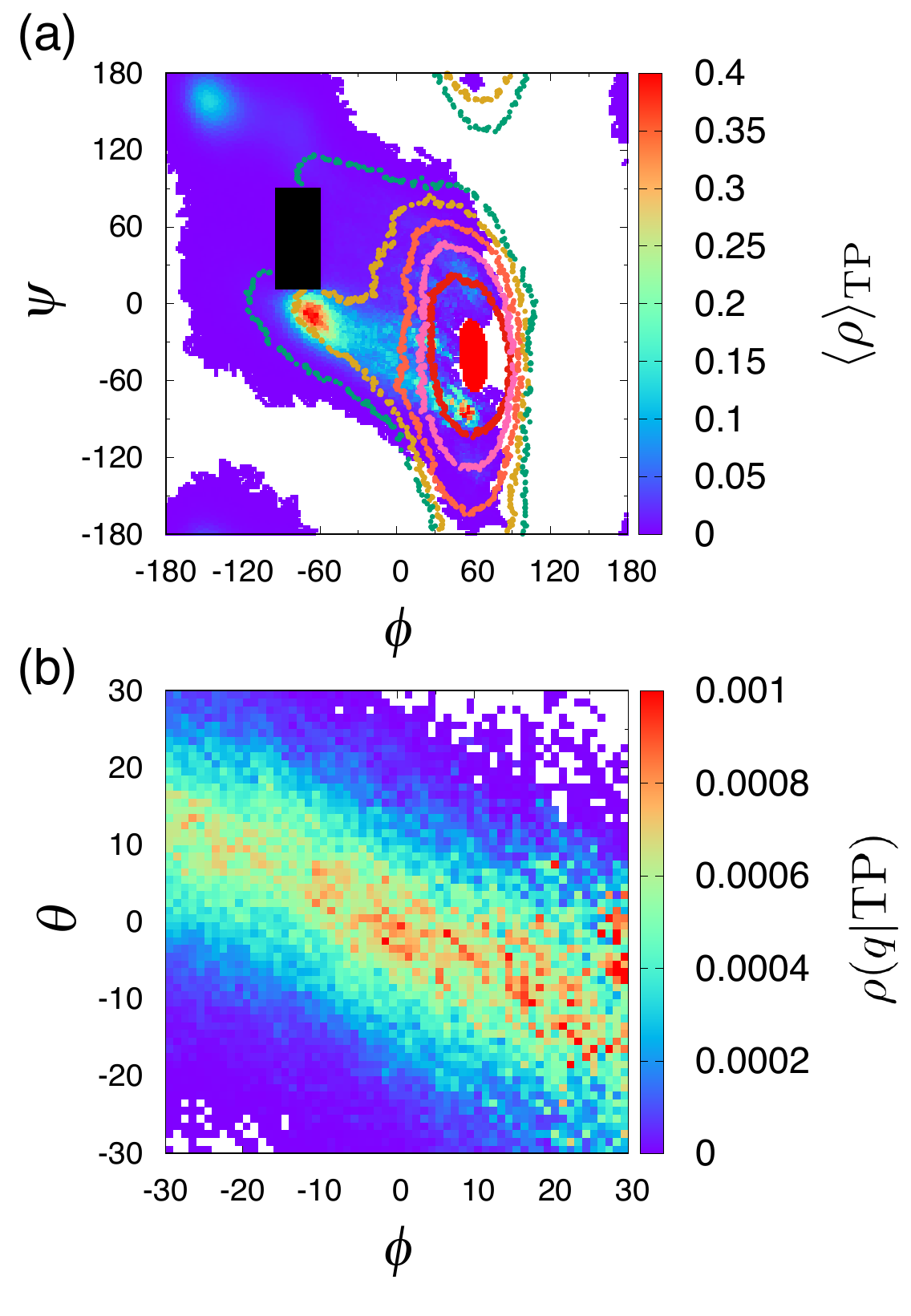}
    \caption{cFFS for alanine dipeptide in vacuum. (a) Initial and
      final states are shown as red and black regions,
      respectively. Configurations collected at $\lambda_0$,
      $\lambda_1$, $\lambda_2$, $\lambda_3$, and $\lambda_4$ are reported as red,
      pink, salmon, gold, and green points, respectively. Color map
      represents the TPE sampling.  $\phi$ and $\psi$ angles are
      reported in degrees. (b) Correlation between $\phi$ and $\theta$
      in the TPE. Color map represents the TPE density of states.}
    \label{fig:adp}
     \vspace{-0.12in}
\end{figure}

In keeping with tradition, we close by demonstrating cFFS on the
C$_{7\text{ax}}$-to-C$_{7\text{eq}}$ conformational change in alanine
dipeptide in vacuum. Details of the simulations and cFFS are reported
in the SI. $\phi$ and $\psi$ backbone dihedral angles were used as CVs
for cFFS. The progression of cFFS is shown in
Fig. \ref{fig:adp}(a). Starting from the C$_{7\text{ax}}$ basin
centered near $\phi=60^\circ$ and $\psi=-30^\circ$, cFFS drives the
system to the C$_{7\text{eq}}$ basin defined by $-94^\circ < \phi <
-60^\circ$ and $12^\circ < \psi < 90^\circ$. The shape of the
interfaces shows that $\phi$ plays the larger role in the transition
and reveals the location of the primary transition tube. The
transition rate constant predicted by cFFS ($k_{AB}^\text{cFFS} = 5.0
\times 10^6$ s$^{-1}$) compares favorably with straightforward
simulation ($k_{AB}^\text{SLD} = 4.8 \times 10^6$ s$^{-1}$). In
Fig. \ref{fig:adp}(b) we show the relationship between $\phi$ and
another dihedral angle, $\theta$, in the TPE. It has been shown that
$\theta$ is part of the reaction
coordinate\cite{Chandler:00:PNAS,Ciccotti:06:JCP}. cFFS captures the
proper relationship between $\phi$ and $\theta$ even though $\theta$
is not one of the CVs used during
cFFS.\cite{Ciccotti:06:JCP,Parrinello:13:PRL}

\section{Discussion}

cFFS helps overcome a few challenges posed by FFS. cFFS allows
  one to try multiple CVs simultaneously. This is beneficial for
  systems where investigators have some a priori insight into the CVs
  that are expected to play a role in the transition, but a detailed
  analysis of the mechanism is missing and the best order parameter
  remains unknown. By using multiple CVs simultaneously and enforcing
  constant flux along an interface, the method can alleviate issues
  associated with poor overlap between $\rho(\lambda^\perp |
  \lambda_i)$ and $P(\lambda_B | \lambda_i; \lambda^\perp)$. Of
  course, it is possible that there are additional important
  orthogonal coordinates beyond the chosen CVs. This situation could
  pose sampling challenges for cFFS. Finally, we demonstrated cFFS
  with a combination of momenta and position based coordinates. This
  may extend the practical applicability of FFS to more ballistic
  systems. FFS depends on stochasticity for trajectory divergence
  between subsequent interfaces, so it will still not be applicable in
  the limit of fully deterministic dynamics.

cFFS can in principle be extended to a large number of
  CVs. However, we surmise the method will not scale well to more than
  three or four CVs. In high dimensional space, the area through which
  trajectories can cross an interface will become exceedingly
  large. From a practical standpoint, this will make it difficult to
  maintain the constant flux condition. From an efficiency standpoint,
  most of each interface will drive the system towards regions of
  phase space which are irrelevant to the transition of interest. Even
  if successful transitions are generated, they will probably
  originate from a tiny subset of the phase points collected at
  $\lambda_0$ and thus be highly correlated. Challenges associated
  with scaling to large numbers of CVs are hardly limited to cFFS. A
  variety of advanced sampling methods, including nonequilibrium
  umbrella sampling\cite{Dinner:07:JCP,Dinner:09:JCP} and
  metadynamics\cite{Parrinello:02:PNAS} have come across similar
  problems. One solution is to collapse the reaction coordinate to a
  single dimension using a string-type
  approach.\cite{Ciccotti:06:JCP,Dinner:09:JCP,Bolhuis:10:JCP} The
  string-type approach will prove difficult to implement in FFS
  without resorting to an iterative scheme requiring multiple FFS
  runs, because each path ensemble in FFS is generated sequentially
  and there is no opportunity to relax the string. Moreover, the
  string-type approach could defeat one of the benefits of cFFS, which
  is that it enables exploration of transitions with multiple tubes.

Extending cFFS to large numbers of CVs will thus require
  alternative approaches. Dimensionality reduction techniques such as
  isomaps\cite{Langford:00:Science,Clementi:06:PNAS} or diffusion
  maps\cite{Zucker:05:PNAS,Debenedetti:11:ChemPhysLett} could be
  employed to reduce a large number of CVs to two or three reduced
  coordinates which capture the largest spread in the data. In this
  manner, multiple transition tubes would hopefully be
  preserved\cite{Clementi:06:PNAS,Debenedetti:15:JCP} within the reduced
  coordinates. Furthermore, several groups are actively working to
  combine machine learning and advanced sampling methods to identify
  important CVs
  on-the-fly.\cite{Kevrekidis:17:PNAS,Ferguson:18:JCC,Tiwary:18:JCP,Pande:18:JCP,Noe:18:JCP}
  We are exploring if such methods or variations thereof can be
  incorporated with cFFS. One challenge to incorporating on-the-fly
  identification of reduced coordinates with FFS-type methods is again
  related to the sequential generation of ensembles. Sampling from the
  initial basin alone is unlikely to reveal reduced coordinates ideal
  for studying the transition. As FFS progresses, sampling from each
  interface ensemble will result in reduced coordinates which
  increasingly describe the transition. However, FFS requires that
  each ensemble be visited sequentially, and changing the definition
  of the reduced coordinates after each ensemble may cause substantial
  difficulty in maintaining this condition.

Studying rare events in simulations is an important and
  challenging problem that has spawned the development of many
  methods in the past decades. Here we restrict our comparison to two
  methods which use multiple CVs to sample, and calculate rate
  constants for rare transitions with unbiased dynamics in equilibrium
  or nonequilibrium systems. Vanden-Eijnden and Venturoli developed a
  method\cite{Venturoli:09:JCP} that calculates the transition rate
  constants and transition paths from the steady state distribution
  under the boundary conditions that state $A$ is a source and state
  $B$ is a sink. The space between the stable states is tiled into
  enclosed Voronoi cells and parallel simulations are performed in
  each cell. The steady state flux and probability distribution can be
  estimated from the time spent in each cell and exchange between
  cells. Like cFFS, the method is applicable to equilibrium as well as
  nonequilibrium systems and does not require that $A$ and $B$ be well
  separated in both variables. Since each parallel path is restricted
  to a single cell, the method may prove advantageous compared with
  cFFS for systems with metastable intermediates. The method does not
  provide direct access to dynamical transition paths, although, in
  principle, transition paths could probably be reconstructed with an
  extensive bookkeeping scheme. It is not immediately apparent which
  method would be better for systems with slowly decorrelating
  transition paths. 

As mentioned in the introduction, Borrreo and
  Escobedo\cite{Escobedo:07:JCP} developed a method to optimize the
  FFS order parameter through a series of FFS runs. The approach uses
  committor information obtained from the prior FFS run to identify
  the best order parameter from a set of specified CVs. The procedure
  can be repeated until TPE sampling or the optimal order parameter
  converges. Like cFFS, the procedure in
  Ref. \citenum{Escobedo:07:JCP} allows FFS to be used in situations
  where there are a number of possible CVs. Since the FFS runs
  themselves are performed along a single order parameter (which may
  be a linear or nonlinear combination of multiple CVs), there is no
  limitation to the number of CVs which can be tested. For certain
  systems this may represent a substantial advantage over cFFS, which
  in current form is practically limited to three or four
  CVs. Unfortunately, the method presented in
  Ref. \citenum{Escobedo:07:JCP} requires multiple (often expensive)
  FFS runs. Additionally, given the sensitivity of FFS sampling to the
  choice of order parameter in the presence of multiple transition
  tubes, we suspect cFFS will perform better for such systems.

Lastly, we would like to comment on the possibility of combining
  a cFFS-type approach with other path sampling methods.  At the most
  basic level, cFFS divides CV space into a fine grid to help define
  regions of phase space and interfaces between those regions with
  arbitrary shape. In cFFS, criteria for boundary identification were
  selected to meet the needs of FFS -- a minimum number of total first
  crossings and constant average flux along the interface to avoid
  biasing the system to proceed in one direction over another. It is
  easy to imagine modifying the boundary identification criteria for
  other applications. Within the family of FFS approaches, it may
  prove fruitful to combine the approach of Borrero and
  Escobedo\cite{Escobedo:07:JCP} with a cFFS-type approach for
  interface definitions. This could allow interfaces with any
  arbitrary shape which could better reproduce the committor
  function. Transition interface sampling is less sensitive to the
  definition of order
  parameter.\cite{vanErp:06:JCP,vanErp:12:AdvChemPhys} However, a
  procedure has been proposed to optimize interface placement given
  the order parameter.\cite{Dellago:11:JCP} This criterion for optimal
  interface placement could be combined with a cFFS-type approach for
  dividing the CV space for transition interface sampling.

\section{Concluding remarks}

We described cFFS, a method to sample rare event transitions along
multiple CVs simultaneously. cFFS uses automated nonlinear interface
placement and reveals on-the-fly the evolution of CVs during a
transition. cFFS was tested with two CVs, but in principle, it can be
extended to three or more. In practice, extending cFFS in current
  form to more than three or four CVs may prove challenging. The
stable states only need to be separated in a combination of CVs, which
may change nonmonotonically between the stable states. We introduced a
criterion of constant flux along each interface to prevent biasing
TPE. cFFS results in correct estimates of the transition rate
constants and TPE sampling on several 2D PESs and the
C$_{7\text{ax}}$-to-C$_{7\text{eq}}$ transition in alanine dipeptide
in vacuum. We additionally demonstrated cFFS on 1D analytical
  potential using one position coordinate and one momenta
  coordinate. cFFS substantially improved upon FFS results on the same
  potential, where only the position coordinate was used an the order
  parameter. On the 2D PESs, cFFS performed particularly well for
systems with hysteresis or multiple transition tubes.  cFFS with
multiple suboptimal order parameters consistently outperformed FFS
with a single suboptimal order parameter. Since optimal order
parameters are not known in most applications of FFS, cFFS with two or
more suboptimal order parameters will be beneficial for studies of
complex systems such as macromolecular conformational transitions and
crystal nucleation.

\begin{acknowledgments}
 We thank the referees for thoughtful suggestions which substantially
 improved an earlier version of this work. This material is based upon
 work supported by the U.S. Department of Energy, Office of Science,
 Office of Basic Energy Sciences, under Award Number
 DE-SC0015448. Clemson University is acknowledged for generous
 allotment of compute time on the Palmetto cluster.
\end{acknowledgments}

\noindent {\bf Supporting Information Available:} Details of Langevin
dynamics, PESs, TPE sampling at $\beta=2.5$, FFS/cFFS sampling on
PES-3 at $\beta=5.0$, and details of alanine dipeptide simulations.

\end{document}


\title{Supporting Information: Contour forward flux sampling:
  Sampling rare events along multiple collective variables}

\author{Ryan S. DeFever}
\author{Sapna Sarupria}
\email{ssarupr@g.clemson.edu}

\affiliation{Department of Chemical \& Biomolecular
  Engineering\\ Clemson University, Clemson, SC 29634}

\maketitle

\vspace{-0.4in}

\section{Details of Langevin dynamics and FFS/cFFS parameters}

The behavior of a particle on the four potential energy surfaces
(PESs) was described by the Langevin equation, $\ddot{\bm{q}} =
-\nabla U(\bm{q}) - \gamma \dot{\bm{q}} + \sqrt{2\gamma k_B T}R(t)$,
where $\bm{q}$ represents the coordinates of the particle, $U(\bm{q})$
is the PES, $\gamma$ is the friction coefficient, $k_B$ is the
Boltzmann constant, $T$ is the reduced temperature, and $R(t)$ is
delta-correlated Gaussian random noise with zero mean and unit
variance. The dynamics were generated with the velocity Verlet
integrator with a time step of 0.01. Simulations were performed at
$\gamma=5.0$. We confirmed that $\gamma=5.0$ provides sufficient
stochasticity for FFS by comparing rates estimated from
straightforward Langevin dynamics (SLD) with rates estimated from FFS
for a range of $\gamma$ from 0.01 to 100 at $\beta=2.5$ (data not
reported). FFS rates agreed with SLD for all surfaces at $\gamma \ge
1.0$. At $\gamma<1.0$ FFS underestimated the rate constant. It is
possible that $\gamma=5.0$ does not provide sufficient stochasticity
at $\beta=5.0$. This may explain why FFS$_\text{opt}$ and cFFS rates
at $\beta=5.0$ agree more closely with SLD for surfaces with a more
flatter and thus more diffusive transition region (PES-1 and PES-4).

SLD results were averaged over 50 and 400-600 independent simulations
of length $10^7$ time units at $\beta=2.5$ and $\beta=5.0$,
respectively. FFS/cFFS results were averaged from three independent
trials. At $\beta=2.5$, FFS/cFFS was performed with 10,000
trajectories per interface at 1,000 configurations per interface. At
$\beta=5.0$, FFS/cFFS was performed with 40,000 trajectories per
interface and 4,000 configurations per interface.

\section{Potential energy surfaces}

Equations for the PESs used in this work are reported in
Eqns. 1--4. Barrier heights are provided in Table
\ref{tab:barriers}. For each surface, the negative-$x$ minimum is
considered state $A$ and the positive-$x$ minimum state $B$. The
potential energy difference from the minimum of $A$ to the lowest
potential energy transition state is 3.4 for all surfaces. PES-1 and
PES-2 have a single transition tube, while PES-3 and PES-4 have two
transition tubes.

\begin{align}
\begin{split}
    V_\text{PES-1}(x,y) = & 0.02(x^4+y^4) - 4\exp(-((x+2)^2 + (y+2)^2)) \\ 
              &- 4\exp(-((x-2)^2 + (y-2)^2)) + 0.3(x-y)^2 + 0.0026
\end{split}
\end{align}

\begin{align}
\begin{split}
    V_\text{PES-2}(x,y) = & 0.02(x^2+y^2)^2 - 5.196\exp(-0.08(x+3.5)^2 - 1.5(y+1.3)^2) \\
              &- 5.196\exp(-0.08(x-3.5)^2 - 1.5(y-1.3)^2) + 0.30914
\end{split}
\end{align}

\begin{align}
\begin{split}
    V_\text{PES-3}(x,y) = & 0.02(x^4+y^4) - 3.73\exp(-((x+2)^2/8 + (y+2)^2/8)) \\
              &- 3.73\exp(-((x-2)^2/8 + (y-2)^2/8)) \\
              &+ 3\exp(-(x^2/2 + y^2/15)) + 2\exp(-(x^2/2 + y^2/2)) - 0.5085
\end{split}
\end{align}

\begin{align}
\begin{split}
    V_\text{PES-4}(x,y) = & -2.93\exp(-((x-3)^2/2+(y-2.5)^2/2)) \\
              &- 2.93\exp(-((x+2)^2/2+(y+2)^2/2)) \\
              &+ 3\exp(-0.32((x+1)^2+(y-2)^2+12(x+y-2.7)^2-1)) \\ 
              &+ 6\exp(-0.15((x-2)^2+(y-1)^2+10(x+y)^2-1)) \\
              &+ 0.005(x^4+y^4) - 0.627
\end{split}
\end{align}

\begin{table}
  \centering
    \caption{Barrier heights for PESs. TS1 and TS2 are the positive-$y$ and
  negative-$y$ transition states, respectively.}
    \label{tab:barriers}
\begin{tabular}{c c c }
\toprule
     & A$\xrightarrow{\mathrm{TS1}}$B & A$\xrightarrow{\mathrm{TS2}}$B  \\
    \midrule
    PES-1   & 3.402  & N/A    \\
    PES-2   & 3.403  & N/A    \\
    PES-3   & 3.403  & 3.403  \\
    PES-4   & 3.405  & 4.143  \\

\bottomrule
\end{tabular}
\end{table}

\section{TPE sampling at $\beta=2.5$}

\begin{figure}
    \centering
    \includegraphics[width=0.9\linewidth]{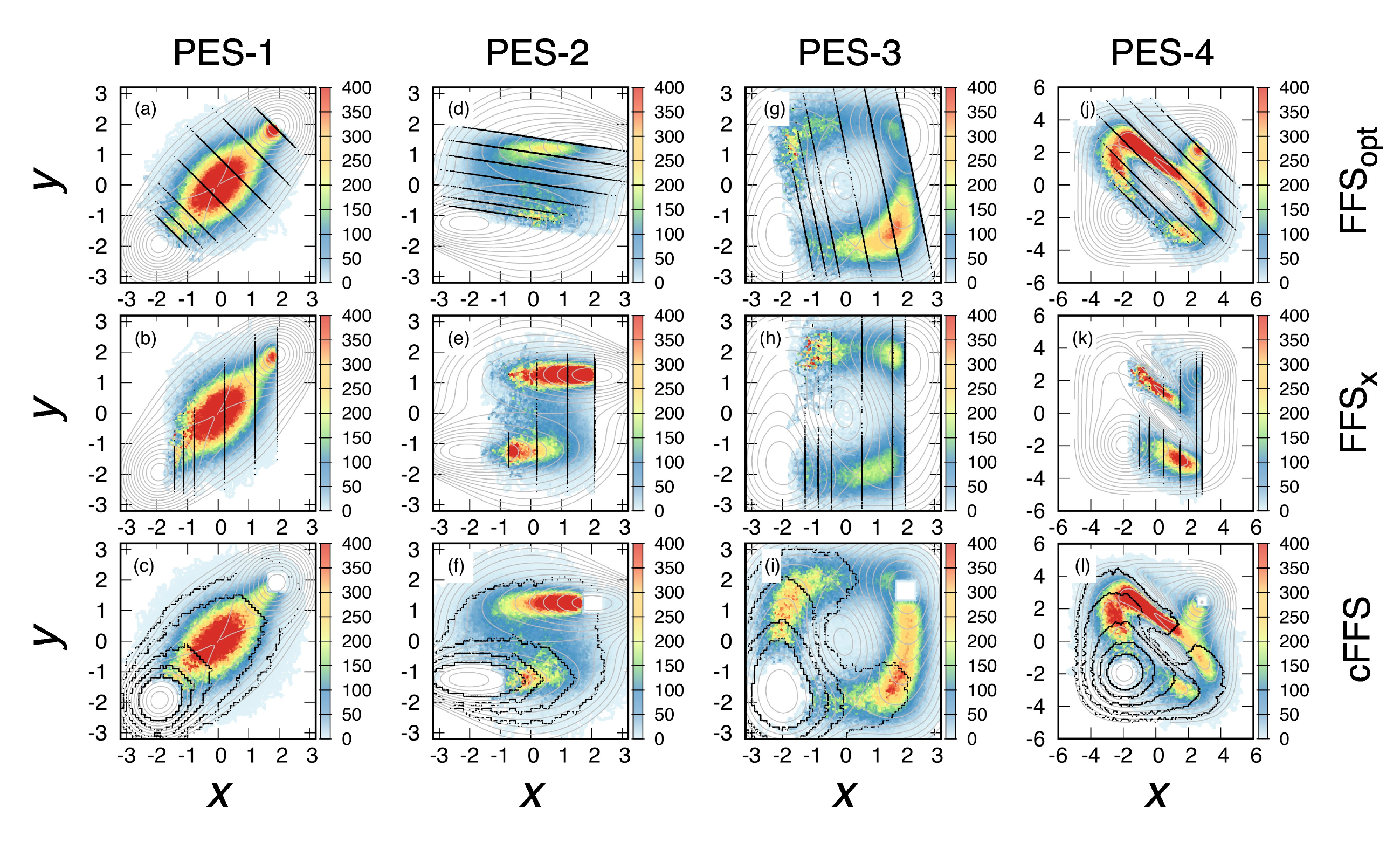}
    \caption{Comparison of interface placement and TPE sampling
      generated with FFS$_\text{opt}$, FFS$_\text{x}$, and cFFS on
      PES-1 -- PES-4 at $\beta=2.5$. PES contours are shown as gray
      lines. Configurations at each interface are shown with black
      points. TPE sampling is represented by the heat map.}
    \label{fig:tpe-b2}
\end{figure}

Transition path ensemble (TPE) sampling at $\beta=2.5$ is reported in
Fig. \ref{fig:tpe-b2}. All methods result in similar sampling to the
SLD results reported in Fig. 1 of the main text. All methods
successfully sample both transition tubes for PES-3 and PES-4 at this
higher temperature.

\section{FFS sampling on PES-3 at $\beta=5.0$}

\begin{figure}
    \centering
    \includegraphics[width=0.6\linewidth]{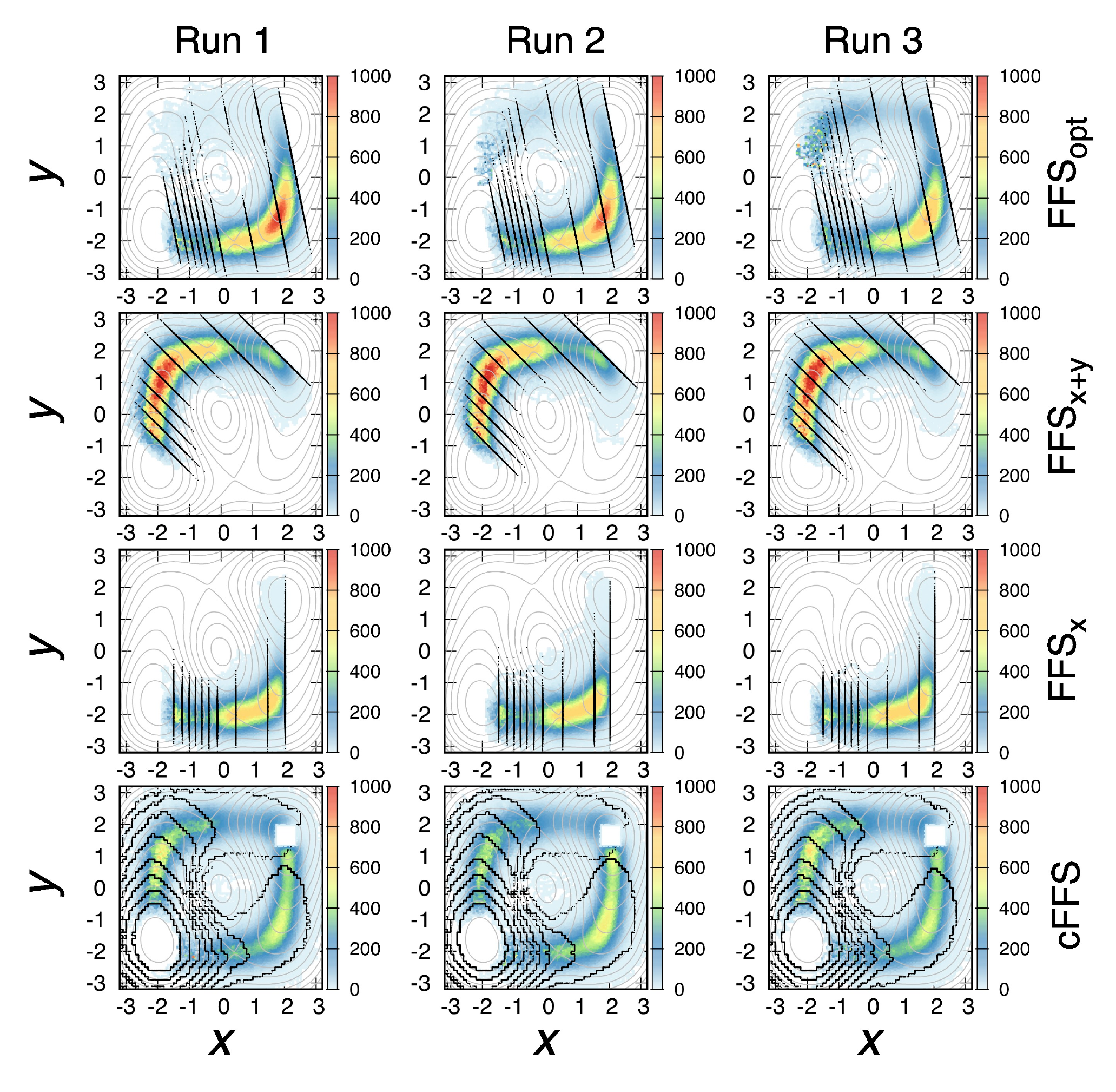}
    \caption{Comparison of interface placement and TPE sampling
      generated with FFS$_\text{opt}$, FFS$_\text{x+y}$,
      FFS$_\text{x}$, and cFFS on PES-3 at $\beta=5.0$. Results are
      shown for all three independent FFS runs for each method. PES
      contours are shown as gray lines. Configurations at each
      interface are shown with black points. TPE sampling is
      represented by the heat map.}
    \label{fig:pes3}
\end{figure}

FFS struggles to sample both transition tubes at $\beta=5.0$ on
PES-3. TPE sampling for all three runs of FFS with three different
order parameters and cFFS are reported in Fig. \ref{fig:pes3}. Even
with the optimal order parameter ($5x+y$), FFS fails to equally sample
both transition tubes. With one suboptimal order parameter ($x+y$)
FFS always samples the positive-$y$ transition tube, while with a
different suboptimal order parameter ($x$), FFS always samples the
negative-$y$ transition tube. cFFS consistently samples both
transition tubes. 

\clearpage

\section{Details of alanine dipeptide simulations}

Alanine dipeptide was simulated in vacuum with Langevin dynamics at
300 K with the leap-frog stochastic dynamics integrator implemented in
GROMACS 2018.\cite{gromacs} The integration time step was 0.002 ps and
$\gamma=100$ ps$^{-1}$. Linear and angular center of mass motion was
removed every step. Alanine dipeptide was represented with the
AMBER99SB force field.\cite{amber99sb} Bonds between heavy atoms and a
hydrogen were constrained with LINCS.\cite{lincs,plincs}

cFFS was tested by investigating the
C$_{7\text{ax}}$-to-C$_{7\text{eq}}$ conformational transition, which
requires surmounting an $\sim$10 $k_BT$
barrier.\cite{Parrinello:18:JPCL} The SLD rate constant was estimated
from 25 independent simulations. Each simulation was initiated from
the C$_{7\text{ax}}$ basin located near $\phi=60^\circ$ and
$\psi=-30^\circ$. Following an energy minimization, the systems were
equilibrated for 1 ns prior to the start of the production runs. Each
production run was continued until the system committed to the
C$_{7\text{eq}}$ basin or for a maximum of 500 ns. 23 of the 25
simulations underwent the conformational transition within 500 ns. The
rate constant was estimated as $k_{AB} = n_{AB}/t_A$ where $n_{AB}$ is
the number of C$_{7\text{ax}}$-to-C$_{7\text{eq}}$ transitions and
$t_A$ is the total simulation time spent in the C$_{7\text{ax}}$
basin, and thus $k_{AB}^\text{SLD} = 4.8 \times 10^6$ s$^{-1}$.

cFFS was performed for the same system. Simulation in basin $A$ was
initiated from an energy minimized configuration in the
C$_{7\text{ax}}$ basin. The system was equilibrated for 1 ns prior to
the start of a 10 ns production simulation. $\phi$ and $\psi$ were
selected as the CVs for cFFS. The grid extended from $-180^\circ$ to
$180^\circ$ in both $\phi$ and $\psi$ with periodic boundaries. A grid
size of 2$^\circ$ was used in both $\phi$ and $\psi$. The bounds of
basin $B$ were defined by examining the free energy landscape reported
in the Supporting Information of
Ref. \citenum{Parrinello:18:JPCL}. The bounds of basin $A$ were
identified with a threshold probability density of $4.0 \times
10^{-4}$. This results in the system spending $\sim$60\% of the time
within the bounds of $A$ during the basin simulation. The flux from
$A$ to $\lambda_0$ was calculated to be $6.15 \times 10^{10}$
s$^{-1}$. Atomic velocities were {\it not} regenerated at the shooting
points as the stochastic dynamics allowed individual trajectories to
diverge. Complete details of the cFFS run are reported in Table
\ref{tab:cffs}. The total probability of reaching $B$ from $\lambda_0$ is $\sum_{i=0}^4 P(\lambda_B | \lambda_i) P(\lambda_i | \lambda_0) = 8.15 \times 10^{-5}$ and
thus $k^{\text{cFFS}}_{AB} = 5.0 \times 10^6$ s$^{-1}$.

\begin{table}[!h]
  \centering
  \caption{Alanine dipeptide cFFS details}
  \label{tab:cffs}
  \vspace{-0.1in}
\setlength{\tabcolsep}{6pt}
\begin{tabular}{ c  c  c  c  c  c  c  c c }
\toprule
$i$ & $N_\text{conf}$ & $N_\text{basin}$ & $N_\text{cross}$ & $N_\text{succ}$ & $N_\text{total}$ & $P(\lambda_{i+1}|\lambda_i)$  & $P(\lambda_i|\lambda_0)$ & $P(\lambda_{B} | \lambda_i )$ \\
\midrule
0   & 615 & 9347 & 653 & 0    & 10000 & 0.0653 & 1.0                  & 0.0      \\	
1   & 653 & 9394 & 606 & 0    &	10000 & 0.0606 & $6.53\times 10^{-2}$ & 0.0      \\	
2   & 606 & 9397 & 586 & 17   &	10000 & 0.0586 & $3.96\times 10^{-3}$ & 0.0017   \\	
3   & 582 & 6471 & 466 & 3063 & 10000 & 0.0466 & $2.32\times 10^{-4}$ & 0.3063   \\	
4   & 466 & 3258 & 0   & 1742 & 5000  & 0.0    & $1.08\times 10^{-5}$ & 0.3484   \\	

\bottomrule
\end{tabular} 
\end{table}